 \documentclass[12pt,preprint]{aastex}

\usepackage{graphicx}
\usepackage{amsmath}   

\title{
    MHD Simulations of Parker Instability Undergoing \\
      Cosmic-Ray Diffusion}
\author{
   Chih-Yueh Wang\thanks{cw5b@msn.com;cywang@phys.cycu.edu.tw}, 
       Ying-Yi Lo\thanks{eddyluo007@pchome.com.tw},
       Chung-Ming Ko\thanks{cmko@astro.ncu.edu.tw}\\
   \affil{$^{1}$Department of Physics, Chung-Yuan Christian University \\
          $^{2}$Institute of Astronomy and Center for Complex Systems,  
       National Central University, Chung-Li, Taiwan 320 \\
    {\sf{Interstellar Medium: New Research, Chapter 5}} }}

\begin{document}
\date{}
\maketitle
\label{firstpage}

\begin{abstract}
Parker instability arises from the presence of magnetic fields in a plasma in a gravitational field such as the interstellar medium (ISM), wherein the magnetic buoyant pressure expels the gas and causes the gas to move along the field lines. The subsequent gravitational collapse of the plasma gas is thought to be responsible for the formation of giant molecular clouds in the Galaxy. The process of mixing in the ISM near the Galactic plane is investigated. The initial ISM is assumed
to consist of two fluids: plasma gas and cosmic-ray particles, in hydrostatic equilibrium, coupled with a uniform, azimuthally-aligned magnetic field. The evolution of the instability is explored in two models: an isothermal exponential-declining density model and a two-layered, hyperbolic tangent temperature model. After a small perturbation, the unstable gas aggregates at the bottom of the magnetic loops and forms dense blobs. The growth rate of the instability decreases as the coupling between the cosmic rays and the plasma becomes stronger (meaning a smaller CR diffusion coefficient). The mixing is enhanced by the cosmic-ray diffusion, while the shape of the condensed gas depends sensitively on the initial equilibrium conditions.
The hyperbolic tangent temperature model produces a more concentrated
and round shape of clumps at the
foot points of rising magnetic arches, like the observed giant molecular cloud,
whereas
the exponential density
model gives rise to a filamentary morphology of the clumpy structure.
When considering a minimum perpendicular or cross field diffusion of cosmic rays,
 which is often substantially smaller than the parallel coefficient $\kappa_{\|}$,
 around $2\%-4\%$ of $\kappa_{\|}$,
the flow speed is significantly increased such that the magnetic loops extend to a greater altitude.
We speculate that the galactic wind flow perpendicular to the galactic disk may be facilitated by Parker instabilities through the cross field diffusion of cosmic rays. 

\end{abstract}


\section{Introduction}

Parker (1966) first identified the instability of a gravitationally stratified gaseous disk in a magneto-hydrostatic equilibrium, like the Galaxy, in response to perturbations due to magnetic buoyancy, for perturbations that occur in the magnetic field lines that lie parallel to the disk plane. This phenomenon is referred to as magnetic buoyancy instability (Hughes~\&~Proctor 1988; Tajima~\&~Shibata~1997)
or Parker instability (Parker 1966; 1979). Solar magnetic activities such as sunspots, which is caused by the emergence of magnetic flux tubes from the interior of the sun into the solar atmosphere (Zwaan 1985, 1987), reflect this instability.  The mushroom shape of the hydrogen cloud GW 123.4-1.5 (Baek,~Kudoh~\&~Tomisaka, 2008) also suggested magnetic flotation and, hence, Parker instability.
In addition, giant dense CO molecular loops close to the Galactic center has been ascribed to the instability (Fukui et al., 2006).
As is expected, gas aggregating at the foot points of the rising magnetic loops eventually collapses, becoming a giant cloud and a region that hosts stellar formations.
On the other hand,
for models galactic dynamos,
Hanasz et al. (2004) found that
if the effect of cosmic rays is incorporated, 
the efficiency of sustaining the galactic magnetic field can be improved. 
Otmianowska-Mazur et al. (2009) suggested that 
the extended, X-shaped magnetic halo structures observed in some edge-on galaxies 
could be attributed to the cosmic-ray driven dynamo.
Actually, based on his instability, Parker (1992) introduced the idea of cosmic-ray driven dynamo to maintain the Galactic magnetic field,
because the cosmic-ray pressure functions just like the magnetic pressure, capable of overpowering the gas pressure inside the magnetic flux tube and making it easier for the gas to rise.

Cosmic rays are a major component in the interstellar medium (ISM) in galaxies, whose energy density is comparable to the kinetic and magnetic energy densities of thermal plasma gas (Ferri\'ere 2001).
As high energy particles endowed with an energy of up to $10^{14-15} \rm~eV$, of which, 90\% are protons,
as long as their gyroradius is significantly smaller than the characteristic spatial scales of the magnetic field, cosmic-ray  particles only propagate along the magnetic field lines.
When the gyroradius is comparable to the scale of magnetic field variation,
cosmic rays interact strongly with the field through
gyroresonant scattering by the magnetic irregularities.
Although the velocity of cosmic-ray particles is close to the speed of light, the bulk motion of cosmic rays is diffusive and the bulk speed is of the order of Alfv\'en speed
(Cesarsky 1980).

The morphology of Parker instability exhibits two modes, i.e. the undular mode 
and the interchange mode. 
The undular mode, also called Parker instability, is excited by perturbations along the magnetic field lines 
(the wavenumber vector $k||$ of the perturbation parallel to the magnetic field $B$),
where the falling gas creates a magnetic buoyancy greater than the restoring magnetic tension. 
The interchange mode, known as flute instability or magnetic Rayleigh-Taylor instability (Kruskal~\&~Schwarzchild~1954), occurs for shorter-wavelength perturbations with $k\bot$ perpendicular to $B$, capable of causing two straight flux tubes to interchange and ultimately reducing the potential energy in the system. The linear growth rate of the interchange mode generally exceeds that of the undular mode owing to a rapid growth of short wavelengths. However, in the nonlinear stage, the undular mode often dominates (Matsumoto~et~al,~1993; Tajima~\&~Shibata~1997). Thus, the undular mode (Parker instability) is more important than the interchange mode in astrophysical problems.

Baierlein (1983) and Matsumoto et al. (1988) performed the first one- and two-dimensional (2D), pure MHD simulations of Parker instability, respectively. In the nonlinear stage, the gas condensates to form giant clouds; in addition, a shock wave appears in the flow along the rising magnetic loop. By applying the simulation results of Matsumoto et al. (1988) to the solar atmosphere, Shibata et al. (1989b; 1990a) demonstrated that the emerging magnetic loop still expands self-similarly during the nonlinar evolution in two dimensions. Kamaya et al. (1996) adopted the supernova explosion as a perturbation in the ISM to trigger nonlinear instability. Earlier, Nozawa (1992) examined the instability deeper in the convectively unstable layer of the solar atmosphere; when considering how magnetic shear affects the flow (Hanawa et al. 1992, Nozawa 2005), although the interchange mode is stabilized, a large unstable thin structure may still arise. Shantanu et al. (1997) and Kim et al. (2000) made a further application of Parker instability to the Galactic disk without cosmic rays.

Parker's original analysis showed that the instability has a maximum growth rate for non-zero $k\bot$, although non-zero $k||$ is the main cause for the instability. In the 3D simulations for both solar and galactic problems given in Matsumoto \& Shibata (1992) and Matsumoto et al. (1993), previous 2D results of cloud formation, presence of shock wave, and self-similar evolution are confirmed. However, the spatial and temporal scale in these studies depends on the $k \bot$. If a larger $k \bot$ is applied, the magnetic loop tends to have a thinner structure and a horizontal expansion, which would suppress the upward expansion (Parker 1979). Similarly thin slice structures have been found in other 3D simulations of 
Kim et al. (1998, 2001, 2001) and Hanasz et al. (2002).

This study describes the evolution of Parker instability undergoing cosmic-ray diffusion using 2D simulations. Effects of the adiabatic index and the initial hydrostatic condition other than an isothermal temperature and a uniform density profile are examined. Two initial profiles are used, i.e., the hyperbolic tangent temperature model, which is often used in the solar atmosphere, and the exponential density model, which is appropriate for Galactic problems.
Various perturbation results are explored.
The physical parameters and initial conditions invoked are appropriate for a galactic ISM in the solar neighborhood.
Section 2--6 details the governing equations, numerical algorithms, normalization,
initial conditions, and grid setup. Section 7 presents the results.
Section 8 draws the conclusion.


\section{Governing Equations}

We investigate the Parker instability with respect to how cosmic rays affect the fluid element in a uniformly-rotating galactic disk, under the influence of external gravity from the galactic center. A local rectangular domain representing the corotating sheet of the  disk in the vertical plane is used for the simulations. The horizontal component of the radially inward external gravity is balanced by the centrifugal force, and hence, in the momentum equation, only the vertical component of external gravity and the Coriolis force are present to account for the rotation. Self-gravity from the plasma gas is not included.

A hydrodynamic approach is adopted, in which the cosmic rays and the thermal plasma are two separate fluid components of a plasma system. The plasma fluid has a mass density of $\rho$, a thermal pressure of $P_{g}$, and a cosmic-ray pressure of $P_c$, all of which are threaded by a frozen-in magnetic field ${\bf B}$. The cosmic-ray energy equation is described based on the diffusion-convection equation
(Drury~\&~V\"olk~1981; Jones~\&~Kang 1990, Ko~1992),
which treats the cosmic rays as a hot massless fluid and neglects the momentum spectrum of cosmic rays to simplify the governing equations. The artificial separation of the cosmic rays from the plasma helps to distinguish the role played by high- and low-energy components.

The cosmic-ray diffusion-convection equation supplements the standard set of ideal MHD equations. The governing equations in the complete vector form are written as
\begin{eqnarray}
   \frac{\partial \rho}{\partial t} &+& {\bf \nabla} \cdot (\rho\textbf{V}) = \textit{0}, \label{parkermass} \\
   \frac{\partial\rho\textbf{V}}{\partial t} &+& {\bf \nabla} \cdot \left[ \rho \textbf{V} \textbf{V}
       +\left( P_{g}+P_{c}+\frac{B^2}{8\pi} \right) \textbf{I} + \frac{\textbf{B}\textbf{B}}{4\pi} \right ] \nonumber \\
   &-& \rho\textbf{g} + 2\rho{\bf \Omega} \times \textbf{V}=0, \label{parkermomentum} \\
   \frac{\partial \textbf{B}}{\partial t} &+&  {\bf \nabla} \times (\textbf{V}\times \textbf{B})=0,\label{parkerinduction}\\
   \frac{\partial}{\partial t}\left(\frac{P_{g}}{\gamma_{g}-1} \right.&+&\left. \frac{1}{2}\rho \it{V}^{2}+\frac{\it{B}^2}{8\pi}\right)
           \nonumber \\
   &+& {\nabla}{\cdot}\left[ \left(\frac{\gamma_{g}}{\gamma_{g}-1} P_{g}+\frac{1}{2}\rho\it{V}^{2}\right)\textbf{V}+
           \frac{c}{4\pi}\textbf{E}{\times}\textbf{B}\right]\nonumber \\
   &+&\textbf{V}{\cdot} ({\nabla}P_{c}-\rho\textbf{g})=0,\label{parkerenerge}\\
   \frac{\partial}{\partial t}\left(\frac{P_c}{\gamma_{c}-1}\right) &+&  {\bf \nabla} \cdot \left(\frac{\gamma_c}{\gamma_c -1}P_{c} \right) \textbf{V}-\textbf{V} \cdot {\bf \nabla} P_{c} \nonumber \\
       &-& {\bf \nabla} \cdot \left[ \overleftrightarrow \kappa  \cdot {\bf \nabla} \left(\frac{P_{c}}{\gamma_{c} -1}\right) \right] =0, \label{parkerCR}
\end{eqnarray}
where ${\bf V}$ denotes the plasma fluid velocity; $\textbf{I}$  denotes a unit tensor; $\gamma_{g}$  denotes the adiabatic index, i.e. ratio of the heat capacity at a constant pressure to that at a constant volume, of the thermal plasma gas; $\gamma_{c}$ refers to the adiabatic index of the cosmic rays; $\overleftrightarrow \kappa$ represents the cosmic-ray diffusivity; ${\bf \Omega}$  represents the rotation angular frequency; and ${\bf g}$ is the external gravitational acceleration. Deriving this equation set involves use of the distribution function in Vlasov equation (Skilling 1975a, b, c). Such governing equations have also been used in the 2D hyperbolic tangent temperature model of Kuwabara, Nakamura~\&~Ko~(2004), and the 3D model of Lo,~Ko~\&~Wang~(2011).

In addition to balancing the energy equations, the term cosmic-ray pressure $P_c$ affects the momentum equation Eq. (\ref{parkermomentum}). However, particles of cosmic rays do not interact with plasma directly; they interact with plasma via the magnetic field. On a microscopic scale, resonant scattering of Alfv\'en waves keeps the cosmic rays nearly isotropically distributed everywhere with respect to the thermal plasma background. A situation in which the cosmic-ray pressure possesses a gradient $\nabla P_c$ influences the motion of the plasma gas. Thus, in formulating the equations, the interaction of cosmic-ray particles with a thermal plasma is represented by the cosmic-ray pressure $P_c$ and its gradient. The transport of the cosmic-ray pressure is described by a macroscopic diffusion coefficient, $\overleftrightarrow \kappa$, an energy-weighted mean diffusion tensor, defined as
\begin{eqnarray}
   \overleftrightarrow{\kappa}  =  (\kappa_{\|}-\kappa_{\bot})\hat{b}\hat{b}+\kappa_{\bot}\delta_{ij}, \label{kappa}
\end{eqnarray}
where $\hat{b}$ denotes a unit vector along the ${\bf B}$ direction.

The limit of this model is mainly that one must assume a priori knowledge of the cosmic-ray pressure and energy density that satisfies the adiabatic index $\gamma_c=1+P_{c}/E_{c}$. Here, Eqs. (\ref{parkermass})--(\ref{parkerCR}) are solved numerically in a local reference frame in Cartesian coordinates, whose center lies at a galactocentric radius $R_{o}$ and orbits the galaxy with a fixed angular velocity $\Omega=\Omega(R_{o})$. In this local frame, the radial, azimuthal, and vertical spatial coordinates are related to the Cartesian coordinates such that
$x=R-R_{o}$, $y=R_{o}\cdot (\phi-\Omega~t)$, and $z=z$.
In two dimensions a slice of flow is simulated, and
so the rotational effect is not present.

\section{Numerical Algorithms}

A finite difference method with operator splitting
is employed to solve the governing equations. Eqs. (\ref{parkermass})--(\ref{parkerenerge})  are in the flux-conservative form and 
solved by 2-Steps Lax-Wendroff explicit method. The cosmic-ray energy equation Eq. (\ref{parkerCR}) is divided into the convection and diffusion parts. The convection part is first converted to a conservative form and then also solved by 2-Steps Lax-Wendroff explicit method. The diffusion part is solved by the Bi-Conjugate Gradients Stabilized (BICGStab) implicit method.

In the 2-step Lax-Wendroff method,
Eqs. (\ref{parkermass})--(\ref{parkerenerge}) and the convection part of Eq. (\ref{parkerCR})
are rewritten in the conservative form:
\begin{eqnarray}
\frac{\partial U}{\partial t} +  \frac{\partial F}{\partial x}+\frac{\partial G}{\partial y}+\frac{\partial H}{\partial z}=S,\label{cfd}
\end{eqnarray}
where  $U$ can be density $\rho$, momentums $\rho{\bf V}$, energy $E_{g}$ and magnetic field $\bf{B}$;
$F$,$G$,$H$ are the flux of $U$ in $x$-,$y$-,$z$- direction; and $S$ is the source term.
The $x$ component of Eq. (\ref{cfd}) is split into two steps:
\begin{eqnarray}
\begin{array}{l}
\left\{
\begin{array}{l}
\left .
\displaystyle \frac{U_{i+1/2}^{n+1/2}-\frac{U_{i}^{n}+U_{i+1}^{n}}{2}}{1/2\Delta t}+
   \frac{F_{i+1}^{n}-F_{i}^{n}}{\Delta x}+S_{i}^{n} =0  \right \} Step1\\[0.6cm]
\left .
\displaystyle \frac{U_{i}^{n+1}-U_{i}^{n}}{\Delta t}+\frac{F_{i+1/2}^{n+1/2}-F_{i-1/2}^{n+1/2}}{\Delta x}+S_{i}^{n+1/2} =0
\right \}Step2.\label{lax-wendroff}
\end{array}
\right .
\end{array}
\end{eqnarray}
where superscript $n$ denotes time advection and subscript $i$ represents space grid.


The diffusion part of Eq. (\ref{parkerCR}) is written as:
$$\frac{\partial E_c}{\partial t}
-{\bf \nabla}\cdot \left[\overleftrightarrow{\kappa}{\bf \nabla}(E_{c}) \right] = 0,$$
whose finite-difference form is expressed as ${\bf A}{\bf x}={\bf b}$,
where the submatrix in ${\bf A}=[{\bf A}_{1}, {\bf A}_{2}, {\bf A}_{3}, \\
 {\bf A}_{4}, {\bf A}_{5}, {\bf A}_{6}, {\bf A}_{7}]$
and the vector ${\bf b}$ is the function of time step $\Delta t$,
the cosmic-ray diffusion coefficient $\overleftrightarrow{\kappa}$,
the magnetic field ${\bf B}$ and the cosmic-ray energy $E_{c}^{n}$ at the $n$th iteration.
The vector ${\bf x}$ is the unknown:
\begin{multline}
{\bf x}  =\Big[E_{c,i,j,k-1}^{n+1}, E_{c,i,j-1,k}^{n+1}, E_{c,i-1,j,k}^{n+1}, E_{c,i,j,k}^{n+1},\\
E_{c,i+1,j,k}^{n+1}, E_{c,i,j+1,k}^{n+1}, E_{c,i,j,k+1}^{n+1} \Big]^{T}. \label{vectorx}
\end{multline}

In order to solve ${\bf x}$ in Eq. (\ref{vectorx}),
the BICGStab method is employed  
to handle asymmetric linear systems and reduce the operations per iteration to $\textit{O}(N^2)$,
where $N$ is the number of unknowns in the discretized domain.
This method is more efficient than the direct solution methods such as $LU$ decomposition,
which require $\textit{O}(N^3)$ operations.
The BICGStab is an iteration method that uses an initial guess ${\bf x}^{0}$ to find a corresponding residual ${\bf r}^{0}={\bf b}-{\bf A}{\bf x}^{0}$,
and then iterate to the $i$-th step ${\bf r}^{i}$ to an accepted value by means of bi-conjugate matrix-vector operation.
For comparison, when solving the nonlinear system $F(y) = 0$ with Newton's method:
\begin{eqnarray}
   y_{i}  =  y_{i-1}-\frac{F(y_{i-1})} {F'(y_{i-1})}  =  y_{i-1}+\delta_{i}, \nonumber
\end{eqnarray}
the correction $\delta_{i}$ is determined by solving the gradient of the function,
whereas in the BICGStab method,
\begin{eqnarray}
\bar{r}  =  r_{0}+\frac{\Vert r_0 \Vert}{\Vert y_{i-1}\Vert}y_{i-1}, \nonumber
\end{eqnarray}
is used for the iteration. 
The BICGStab treatment for cosmic-ray energy diffusion equation is an implicit method, and thus
the Courant condition is not affected.

\section{Normalization and Initial Values}

The parameters and variables used in the governing equations withstand an extremely large contrast when expressed in dimensional units, possibly incurring
significant errors due to the presence of ill-conditioned matrices and thus infeasible for numerical calculations. To overcome such numerical difficulties, the above equations are normalized to non-dimensional values that are close to unity. The non-dimensional values is denoted by the superscript `*' while the scaling factors are denoted by a `0' subscript. Some of the scaling factors also represent the quantities in equilibrium at the midplane of the Galactic disk.

The length variables are scaled based on the pressure scale height $H_0$:
$x^{*}=x/H_{0}$, $y^{*}=y/H_{0}$, and $z^{*}=z/H_{0}$. $H_{0}$ is determined by integrating the hydrostatic equilibrium equation:
\begin{eqnarray}
\frac{d}{dz}\left[\left(1+\frac{1}{\alpha}+\frac{1}{\beta}\right)P_{g}\right] + \rho g_{z}=0,\label{hydrostatic01}
\end{eqnarray}
\begin{equation}
-\int\limits_{P_{g0}}^{P_{g0}/e}\frac{dP_{g}}{P_{g}}=\int\limits_{0}^{H_{0}}\frac{\gamma_{g}~g_{z}}{C_{s0}^{2}(1+1/\alpha+1/\beta)}dz,
\end{equation}
given the ideal gas law,
the values of sound speed $C_{s0}$, the gravitational acceleration $g_{z}$ in the $z$-direction, the ratio of plasma pressure to magnetic pressure $\alpha$, the ratio of plasma pressure to cosmic-ray pressure $\beta$; and $\gamma_g$.
The gas at the Galactic disk plane is nearly isothermal such that $\gamma_{g}\approx 1.0$. The sound speed at the disk plane is $C_{s0}=10 \rm{~km/s} $.
Thus, for $\alpha=\beta=1$ and
$g_{z}=2.28\times10^{-8} \rm{~cm~s^{-2}}$,
$H_0 = 50 \rm{~pc} = 1.54 \times 10^{20} \rm{~cm}$.
Given $C_{s0}$, the velocity is normalized to $V^{*}=V/C_{s0}$,
and the time scaling factor $\tau_{0}$ is defined as $H_{0}/C_{s0}$, i.e. $\tau_{0}=1.54 \times 10^{14} \rm{~sec} \approx 5 \rm{~Myr}.$
The ISM density is used to normalize the density, with $\rho_{0}$ being about one atom per cubic cm in the ISM,
or $\rho_{0}=1.6 \times 10^{-24} \rm{~g~cm^{-3}}$.

With the above scaling parameters, the mass equation Eq. (\ref{parkermass}) becomes
\begin{eqnarray}
\partial(\rho_{0}\rho^{*})/\partial(\tau_{0}t^{*})+ 1/H_{0}\nabla^{*}\cdot(\rho_{0}\rho^{*}C_{s0}\bf{V}^{*})=0, \nonumber
\end{eqnarray}
%
which can be converted into
$\partial \rho^{*}/\partial t^{*}+\nabla^{*}\cdot(\rho^{*}\bf{V}^{*})=0$. 
Similarly,
the factor $\rho_{0}C_{s0}/(H_{0}/C_{s0})=\rho_{0}C_{s0}^{2}/H_{0}$ at the left hand side of the momentum equation Eq. (\ref{parkermomentum}) is balanced by the same factor at the right hand side.
The normalized plasma gas pressure is
$P_{g}^{*}=P_{g}/P_{g0}$, where
$P_{g0}=C_{s0}^{2}\rho_{0}=1.6\times~10^{-12} \rm{~g~cm^{-2}~s^{-2}}$,
and so the gradient of plasma pressure becomes
$\rho_{0}C_{s0}^{2} \nabla^{*}P_{g}^{*} /H_{0}$.

Scaling of the cosmic-ray pressure also employs the gas pressure, $P_{c}^{*}=P_{c}/P_{g0}$.
The scaling factor for the magnetic field, $B_{0}$, is determined by
$B_{0}^2 = P_{g0}$, and for the Galaxy,
$B_{0}=\sqrt{P_{g0}}=1.26 \times 10^{-6} \rm{~Gauss}$.

The gravitational acceleration is scaled to
$g_{0}=C_{s0}^{2}/H_{0} =6.49\times 10^{-9} \rm{~cm~s^{-2}}$,
%
while the scaling factor for the rotating angular frequency is the reverse of time:
$\Omega_{0}=C_{s0}/H_{0}=0.65\times 10^{-26} \rm~s^{-1}$.
Using the same scaling parameters,
the momentum equation Eq. (\ref{parkermomentum}),
the induction equation Eq. (\ref{parkerinduction}) and energy equation Eq. (\ref{parkerenerge}) can all be converted to dimensionless values.

Finally, in Eq. (\ref{parkerCR}),
the cosmic-ray diffusion coefficient $\kappa$ is normalized to $\kappa^{*}=\kappa/\kappa_{0}$, where
$\kappa_{0}=C_{s0}H_{0}=1.54\times~10^{24} \rm{~cm^{2}~s^{-1}}$.
Theoretical calculations using path length distribution, life-time of radioactive secondary CR, etc.
suggested that
the diffusion coefficient for CR above 1 GeV in the Milky Way Galaxy is
$1-3\times 10^{28} \rm{~cm^{2}~s^{-1}}$
(Berezinskii et al. 1990; Ptuskin 2001).
Due to numerical difficulties with strong magnetic field and large diffusion coefficient,
this work adopts a magnetic field strength of several $\mu G$,
like those in the solar vicinity,
and a $\sim 10-100$ times lower $\kappa$ than suggested.
Crocker et al. (2010) suggest a lower limit of ∼ 50 $\mu G$ on 400
pc scales near the Galactic center.
However, measurements of the amplitude of the magnetic field
depend on the spatial scale and the energy equipartition or pressure equilibrium among various Galactic
components,
which may differ by 2 orders of magnitude.
Pure MHD models with
strong magnetic field have been presented in Machida et al.
(2009) and Takahashi et al. (2009).

\section{Initial Equilibrium Background}

We adopt two initial equilibrium backgrounds: a temperature distribution that follows a hyperbolic tangent profile and a density distribution that follows an exponential profile. Both these backgrounds bear a declining density profile with height. They
differ mainly in that the hyperbolic tangent temperature has an ascending transition zone that divides the distribution into two distinct regions, while the exponential density follows a continuous decline.
Given the temperature or density profile, the profiles for the density/temperature, pressure, magnetic field and other variables can be derived from the hydrostatic condition.

In the isothermal case the criteria for Parker instability to develop is $\frac{d}{dz}(B/\rho)<0$. When the rising field lines grow, the flow becomes unstable. In our cases, the growth of the instability depends on additional parameters such as the width of transition region and the height of Galactic halo (disk thickness).

\subsection{Hyperbolic Tangent Temperature}

The hyperbolic tangent temperature profile is a two-temperature, layered disk (Shibata et al. 1989a), described by
\begin{eqnarray}
C_{s}^{2}(z) = T(z) = T_{0} + (T_{h}-T_{0})\,\frac{1}{2}\left[\tanh\left(\frac{z-z_{h}}{w_{tr}}\right)+1\right],
\label{2-layer-temp}
\end{eqnarray}
where $C_{s}(z)$ is the sound speed, $T_{0}=10^{4} \rm{~K}$ is the disk temperature and $T_{h}=25T_{0}$ is the halo temperature.
The initial dimensionless temperature is 1.0,  equivalent to $10^{4} \rm{~K}$. Given the ideal gas law $P_g=\rho T /\gamma_{g}$
and the gravitational acceleration $g_{z}$, the initial density profile is solved using the hydrostatic equation 
 Eq. (\ref{hydrostatic01}).
A constant acceleration $g_z$ is assumed because
the grid domain is small and not far way from the disk plane.
The dimensionless gravitational acceleration is set as
$1.0/\gamma_{g}$ (dimensionless units),
equivalent to
$6.49/\gamma_{g}\times 10^{-9} \rm{~cm~s^{-2}}$,
or $\sim 6.18 \times 10^{-9} \rm{~cm~s^{-2}}$, for $\gamma_{g}=1.05$.
Acceleration can vary widely from galaxy to galaxy.
Although a little higher, this value is comparable to the value derived from observations of
the spatial density distribution and the velocity-distance correlation
$500$ pc above the Galactic disk midplane,
  $4\times 10^{-9} \rm{~cm~s^{-2}}$ (Oort 1960; Bahcall 1984; Kuijken \& Gilmore 1989).
The initial density at the disk plane is $1.0$ (dimensionless units), equivalent to $1.6 \times 10^{-24} \rm{~g~cm^{-3}}$.

After obtaining the density profile,
the plasma pressure profile $P_{g} = C_{s}^{2}\rho/\gamma_{g}$ is derived.
The magnetic field is assumed to align in the $x$-direction and vary with height $z$,
and $B_{x}(z)=\sqrt{8\pi P_{g}/\alpha}$.
The initial dimensionless pressure at the disk is $1.0$ 
($2\times 10^{-12} \rm~g~cm^{-2}~s^{-2}$).
The initial magnetic field at the disk for $\alpha=1$ is $\sqrt{8\pi}\sim 5$ (dimensionless units),
or $5B_{0}=6.34$ ($\mu \rm Gauss$),
This field strength is very close to the radio synchrotron measurement of 6 $\mu \rm Gauss$
averaged over a radius of about 1 kpc around the Sun (Beck 2009).

The initial cosmic-ray pressure at disk for $\beta=1$ is $1.0$ (dimensionless units),
or $2\times 10^{-12} \rm~g~cm^{-1}~s^{-2}$.
The rotating angular frequency $\Omega$ (not used in this work) is a free parameter;
$\Omega^{*}=1$
gives
$0.65\times 10^{-26} \rm~s^{-1}$,
or approximately $7$ times the angular frequency at our Sun,
$\Omega_{\odot}=220/7.6 \rm~km~s^{-1}~kpc^{-1}$.
The initial equilibrium background is assumed quiescent.  No initial $y$-velocity and
no rotation effects are present in the system.

\subsection{Exponential density}

The exponential density case produces a nearly isothermal background that is similar to the isothermal model of Kim et al. (1998). 
However, since $\gamma_{g}=1.0$ imposes singularity in the energy equation,
$\gamma_{g} > 1 $ is adopted.
The exponential density profile is defined as:
\begin{eqnarray}
  \rho =\rho_{0}\exp\left(-\frac{z}{H_{\rho}}\right),\label{iso-rho}
\end{eqnarray}
where $\rho_{0}=1.0$ (dimensionless units), equivalent to
$1.6\times 10^{-24} \rm~g~cm^{-3}$;
the density scale $H_{\rho}=4.0$ (dimensionless units) is equivalent to
$6.16\times 10^{20} \rm~cm$.
The plasma pressure follows $P_{g}=\rho T/{\gamma_{g}}$.
Using Eq. (\ref{hydrostatic01}), 
the temperature becomes:
\begin{eqnarray}
T_{0} = \frac{H_{\rho}}{(1+\frac{1}{\alpha}+\frac{1}{\beta})}\,\frac{\gamma_{g}}{g_{z}}, \label{iso-temp}
\end{eqnarray}
which is a constant. A minimum value of the adiabatic index $\gamma_{g}=1.01$ is adopted in this model.
Similar to the hyperbolic tangent model, the initial magnetic field
and cosmic-ray pressure are defined through the parameter $\alpha$ and $\beta$,
$B_{x}(z)=\sqrt{8\pi P_{g}/\alpha}$
and
$P_{c}=P_{g}/\beta$.

\section{Grid Setup and Perturbation}

2D simulations are performed in a rectangular domain in the $x-z$ plane.
To excite the instability,
velocity perturbations are added to the quiescent background.
Two perturbation forms are examined: eigenmode sinusoidal wave and random seed.
The strong explosive perturbation in CR pressure developed from supernova explosions has been considered in Kuwabara et al. (2004).
Notably, although supernova shocks has been believed as efficient cosmic ray accelerator,
channeling at least 10\% of the supernova explosion energy into cosmic rays,
signature of cosmic-ray protons,
the key observational evidence in support of the claim that isolated supernova remnants (SNRs) are the main accelerators of galactic CRs, has remained elusive after decades of
observational effort
(Allen, Houck, \& Sturner 2009).
Calculations of the maximum energies of shock-accelerated electrons 
(Reynolds \& Keohane 1999)
and hydrodynamic simulations of Rayleigh-Taylor instabilities in SNRs
(Ferrand et al. 2010, Fraschetti et al. 2010, Wang 2011)
both indicate
that young SNRs could not be responsible for the highest-energy Galactic cosmic rays, 
unless an unrealistically high injection rate of cosmic rays,
and thus an enormous energy loss in SNRs, is invoked.
Precise measurements of cosmic-ray proton and helium spectra by Adriani et al. (2011)
furher excludes 
a single power law energy distribution 
for the two species
and thus challenged the supernova acceleration paradigm.

The sinusoidal perturbation employs a perturbing velocity $V_x$ described~by
\begin{eqnarray}
V_x = 0.05~C_{s0} \sin\left(\frac{2\pi(x-x_{0})}{\lambda}\right),\nonumber
\end{eqnarray}
where $\lambda=20H_0$ is the most unstable wavelength derived from the linear analysis
for $\kappa_{\|}=200$ (Kuwabara et al. 2004).
$V_x$ is applied within the region of $4H_0<z<8H_0$ and
$|x-x_0|<1/2\lambda$, where $x_{0}=40$ (dimensionless unit).
The grid consists of
$102\times~301$ zones
in a rectangular region of $4 ~\mbox{kpc} \times 3.5~\mbox{kpc}$.
The horizontal length of each grid zone is
$\Delta x=0.8H_{0}$,
while the vertical length is nonuniform:
\begin{eqnarray}
\Delta z = \left\{
\begin{array}{cl}
0.15 H_{0}, & \mbox{$0\leq z < 25H_{0}$} \\
min\{1.05\Delta z, \Delta z_{max}\}, & \mbox{otherwise}
\end{array},\nonumber
\right.
\end{eqnarray}
$\Delta z$ is increased above $z=25H_0$ by a ratio of 1.05, until $\Delta z_{max}=5\Delta~z~(at~z=0)$.
The upper grid domain accommodates the unperturbed magnetic field
to avoid outflows across the grid boundary and spurious results.

For the random velocity perturbation (Shore \& Larosa 1999),
a series of random velocities $V_{x}$ and $V_{z}$ with a maximum amplitude of $5\%$
is added horizontally into the background (Chou et al. 2000).
The grid covers an area of
$9~\mbox{kpc} \times 11.5~\mbox{kpc}$ using
$512\times~512$ zones,
with
$\Delta x=0.35H_{0}$ and
\begin{eqnarray}
\Delta z = \left\{
\begin{array}{cl}
0.15 H_{0}, & \mbox{$0\leq z < 35H_{0}$} \\
min\{1.05\Delta z, \Delta z_{max}\}, & \mbox{otherwise}
\end{array}.\nonumber
\right.
\end{eqnarray}



\section{Results}

\subsection{Pure MHD Models}

We examine how cosmic-ray diffusion affects Parker instability by comparing a series of pure MHD simulations with simulations that include the cosmic-ray effect. Figure~1 illustrates those results using sinusoidal eigenmode perturbations. Figure~1(a) 
presents the hyperbolic tangent temperature model, while Figure~1(b) 
displays the exponential density model at the same epoch. In these figures a grid domain of $z\leq 60$ that just lies within the maximum height of the magnetic loops is shown. Logarithmic values of variables in a dimensionless unit are presented in color contours. White solid lines represent the magnetic field lines, and small white arrows refer to the velocity vectors. The superscript `$*$' denoting dimensionless values is omitted hereafter. 
The exponential model shows a faster growth of the instability because the decline in the gas 
density facilitates the instability.

\begin{figure} 
\centering
 {
    \includegraphics[width=2in]{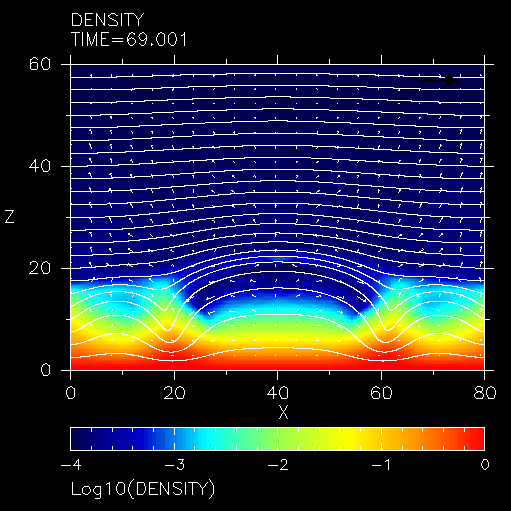}
    \label{f01a} \;\;\;
  }
{\includegraphics[width=2in]{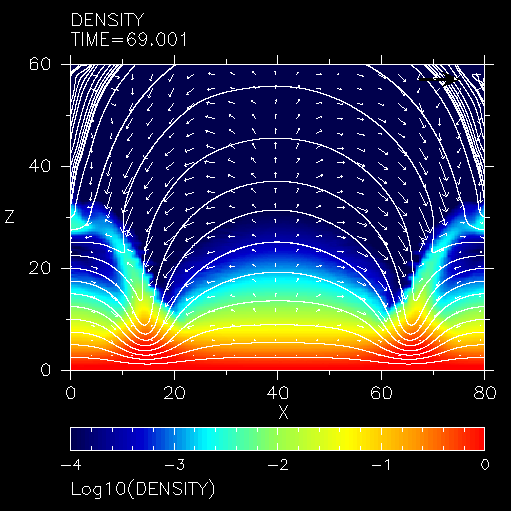}   \label{f01b}  }
\caption{Logarithmic density contours in the pure MHD case developed from a sinusoidal perturbation with $\alpha=1$ at $t\sim~69$. (a) A case of hyperbolic tangent temperature; (b) a case of exponential density.}
\end{figure}

\subsection{Cosmic-Ray Diffusion Coefficient}

Figure~2 compares the maximum height of the magnetic loops versus the  cosmic-ray diffusion coefficient $\kappa_\|$ in both models at $t=33$, when the numerical time step becomes very restrained.
A smaller CR diffusion coefficient means stronger coupling between the cosmic rays and the plasma. 
A maximum of $\kappa_{\|}=200$ ($1.54\times~10^{24} \rm~cm^{2}~s^{-1})$ is examined.
In both models, the instability becomes more prominent with an increasing $\kappa_\|$, and the mixing extends to a similarly maximum height of $z\sim55$. Because the gradient of cosmic-ray pressure $\nabla P_{c}$ declines with an increasing diffusivity, cosmic-ray diffusion facilitates flow instability. Kuwabara et al. (2004) studied the instability in the hyperbolic tangent temperature case. 
For small-$\kappa_{\|}$,                                                                            the instability is impeded by the CR pressure gradient force that interferes with the falling motion of the matter,                                                                                     while for large-$\kappa_{\|}$, the magnetic loop can rise to larger scales.   
Our simulations reach a similar outcome.
In Kuwabara et al (2004), 
the case of strong explosive perturbation in CR pressure presumably from supernova 
yielded
 a larger growth rate for a smaller diffusion coefficient, but such a reversed correlation only occurs 
in the early stage;
  in the later stage the growth rate becomes smaller when compared to that of a large diffusion 
  coefficient model.
In the 3D models of Lo,~Ko~\&~Wang~(2011) employing sinusoidal perturbations, however, the instability growth 
first rises with $\kappa_{\|}$  
but then drops when $\kappa_{\|} \ge 32$,
because under strong diffusion, 
the falling of the gas becomes supersonic; as a result of shock,
the cosmic rays are redistributed with a smoother pressure gradient, 
which tends to stabilize the flow.


\begin{figure}[h] 
\centering
   \includegraphics[width=4in]{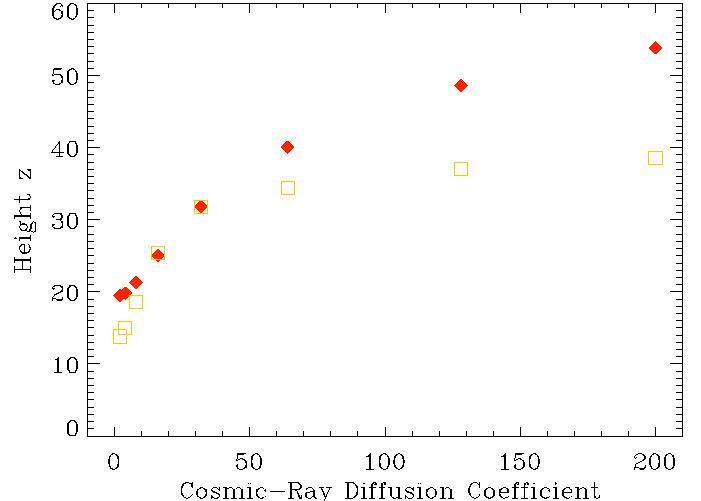}
  \label{f02}
\caption{Maximum height of loops $z$ versus cosmic-ray diffusion coefficient $\kappa_{\|}$ at $t \sim 33$. Diamond denotes the exponential density model and square denote the hyperbolic tangent temperature model.}
\end{figure}

%

\subsection{Parameter $\alpha$ and $\beta$}

The scale height $H_0 = 50 \rm~pc$ is set for $\alpha=\beta=1$. Varying $\alpha=P_{g}/P_{B}$ modifies the length scale, resulting in different flow structures.
Figure~3 presents snapshots of the hyperbolic tangent temperature model with a tenfold gas pressure,
$\alpha=10.0$; 
the flow extends to $z\sim 40$,
which displays more prominent mixing than the $\alpha=1$ case ($z\sim 20$). Other parameters used are $\beta=1.0$; $\gamma_{g}=1.05$; $\gamma_{c}=4/3$;
$w_{tr}=0.6$
(dimensionless units, $30 \rm{~pc}$ or $0.92\times~10^{20} \rm{~cm}$; $z_{h}=18$ ($900 \rm{~pc}$ or
$1.39\times~10^{23} \rm{~cm}$; $\kappa_{\|}=2.0$
($1.54\times 10^{23} \rm{~cm^{2}~s^{-1}}$); and $\kappa_{\perp}=0.0$.

\begin{figure}
\begin{center}
\includegraphics[width=2in]
       {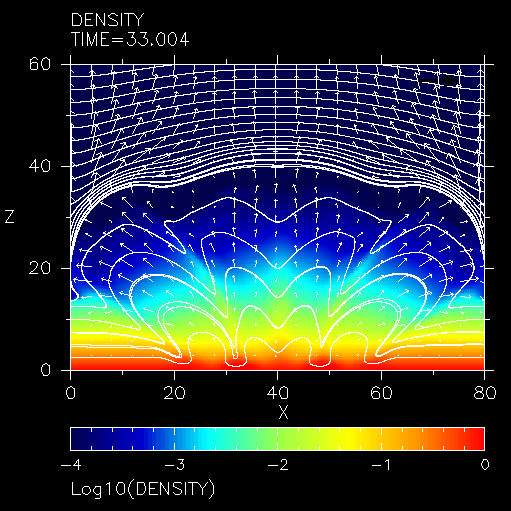} \;\;\;\;
\label{f04a}
\includegraphics
   [width=2in]{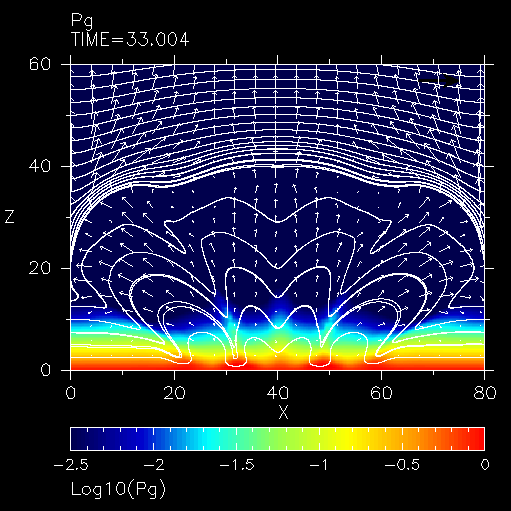}
\label{f04b} \\[0.2cm]
\centering\includegraphics
   [width=2in]{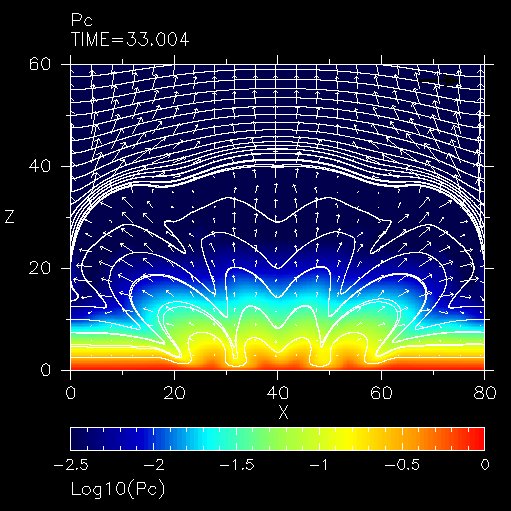}
\label{f04c}
\end{center}
\caption{Snapshots of logarithmic values of density, pressure, and cosmic-ray pressure contours for hyperbolic tangent temperature model at $t\sim~33$ with $\alpha=10$ and $\kappa_{\bot}=0$.} 
\end{figure}

\subsection{Parameter $\gamma_g$ and $w_{tr}$}

The temperature distribution is assumed to be uniform in the exponential density model. In this case, modifying $\alpha$ and $\beta$  changes only the isothermal temperature value. Notably, the parameter $\beta=P_{g}/P_{c}$ should also influence the scale length (see the scaling for $H_0$). However, because cosmic rays largely diffuse along the magnetic field lines, varying $\beta$ does not significantly affect the scale height.

The isothermal case is approximated by using a minimum adiabatic index of $\gamma_{g}=1.01$. In an ISM, $\gamma_{g}$ should be significantly smaller than the ideal value $5/3$ and close to $1.0$, because thermal instability smooths out the temperature gradient. However, because the diffusion of cosmic rays along the magnetic field lines is only slight and the instability growth expedites for more condensed gas, thermal instability is expected to be deterred when cosmic-ray diffusion is present (Shadmehri 2009). This phenomenon may resemble the effect of increasing $\gamma_{g}$ above the typical value $1.05$ that is used in most of the simulations for Parker instability. According to our results, instability is dampened as $\gamma$ increases; with $\gamma=5/3$, the flow remains quiescent at $t\sim 98$ (Figure~4a). 

In the hyperbolic tangent temperature model, an increasing width of transition region $w_{tr}$ makes the instability less feasible (Figure~4b). 
Suppression of the instability is attributed to the rapid rise of temperature and the flattening of density near the transition region.

\begin{figure}[h] 
\begin{center}
      \includegraphics[width=3in]{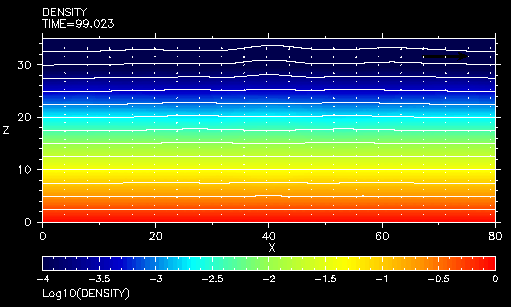}
 \label{f05b}  \\[0.2cm]
  \includegraphics[width=3in]{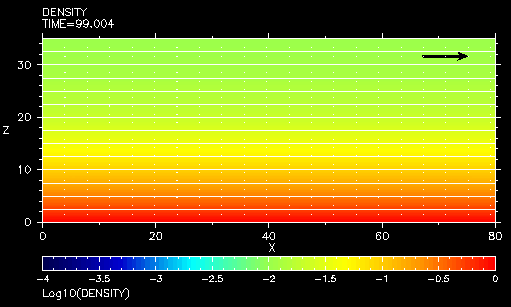}
 \label{f05a}
\end{center}
\caption{Density contours with $\kappa_{\bot}=0$ for (a) $\gamma_g=5/3$ and (b) $w_{tr}=10.0$ at $t\sim 98$ in the hyperbolic tangent model.
         Figure~(a) shows small fluctuations while Figure~(b) does not.}
\end{figure}

\subsection{Random Perturbation}
The case using random velocity perturbation may reflect the circumstances closer to the actual situation in an ISM. In this case, the development of instability is slower
than the case using eigenmode perturbation,
Figure~5  reveals that the exponential density model exhibits slightly higher floatation of the magnetic loops, but similarly strong mixing and filamentary structures are induced in both models.

\begin{figure}[h] 
\begin{center}
      \includegraphics[width=3in]{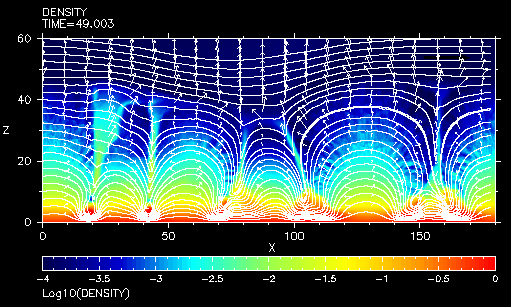}
 \label{f06a}
 \\[0.2cm]
    \includegraphics[width=3in]{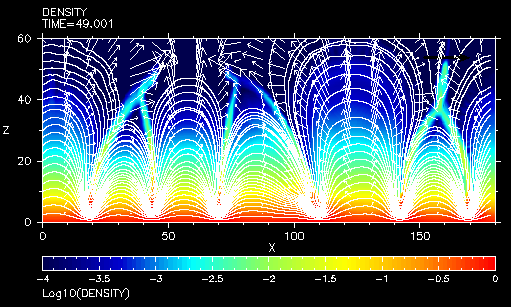}
  \label{f06b}
 \end{center}
\caption{Density contours with $\kappa_{\bot}=0$ at $t\sim~49$ for (a) hyperbolic tangent temperature (b) exponential density, using random perturbation.}
\end{figure}

\subsection{Trebly Sinusoidal Perturbation with $\kappa_{\bot}$}

The perpendicular or cross field lines diffusion coefficient $\kappa_{\bot}$ (see Eq. \ref{kappa}) is often substantially smaller than the parallel coefficient $\kappa_{\|}$, which is only around $2\%-4\%$ of $\kappa_{\|}$ (Ryu et al. 2003). 
%
%
We found that when incorporating cosmic-ray cross field line diffusion $\kappa_{\bot}$, the magnetic field lines are aligned more vertically to the disk. As such a process may be related to the acceleration of galactic wind, a new initial condition using the same equilibrium background
is invoked to examine the case involving $\kappa_{\bot}$.

A grid of $512\times~512$ zones in a domain of
$9 \mbox{kpc} \times 11.5 \mbox{kpc}$ is adopted,
with each zone having a size identical to the case of random perturbation.  The applied velocity perturbation is similar to our previous sinusoidal eigen mode perturbation, but within a finite rectangular region of $|x-x_0|<3/2\lambda$, where $x_0=90.0$, along with other similar parameters: $w_{tr}=0.6$; $z_{h}=18$; $\alpha=\beta=1$; $\gamma_{g}=1.05$; $\gamma_{c}=4/3$; $\kappa_{\|}=2.0$; and $\kappa_{\perp}=0.02$.  

Figure~6 shows the exponential case at the dimensionless time $t\sim 30~\rm~(150~Myr)$.
Three equally-sized loops arise and extend to $z\ge50$.
The flow pattern resembles the hyperbolic tangent case in Kuwabara et al. (2004).
The interaction between loops becomes more pronounced at later times; the central one becomes compressed by two outer adjacent loops, while
 the loops continually rise to a height of $2.0~\mbox{kpc}$. At $t=60$, the flow reaches a maximum vertical height and becomes saturated.
Different from the case without $\kappa_{\bot}$, the flow velocity is significantly increased
such that the loops rise to a much higher altitude, $z\ge 80$.
Therefore, even at a value of $1\%$ of $\kappa_{\|}$,   
the effect of $\kappa_{\bot}$ on the mixing appears significant (Figure~6). 
Although such a result contradicts the estimate made by linear analysis  (Ryu et al. 2003),
it indicates that Parker instability is likely to contribute to the galactic wind acceleration perpendicular to the galactic disk (Wiegelmann, Schindler, \& Neukirch
 1997).

\begin{figure} 
\centering
   \includegraphics[scale=0.45]{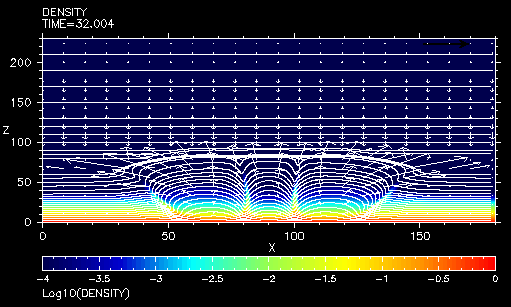}
  \label{f7}
\caption{Density contour of the exponential density model at $t\sim~32$ using trebly sinusoidal perturbation and a cosmic-ray cross field line diffusion $\kappa_{\bot} = 2\%$.
The high velocity indicates Parker instability may contribute to the galactic winds.}
\end{figure}



\section{Conclusion}

This study has elucidated the development of Parker instability undergoing cosmic-ray diffusion using 2D MHD simulations. Exactly how the initial conditions affect the growth of instability is examined.
Two equilibrium backgrounds are constructed based on hyperbolic tangent temperature and exponential density,
with variations in the physical parameters and perturbations.

Our simulations 
gives rise to a structure
that is closely resembles the long, thin wave-like sheared helmet plumes in the solar corona, caused by magnetic buoyancy (Shore \& Larosa 1999). The perpendicular component of cosmic-ray diffusion coefficient $\kappa_{\bot}$ is included as well. When $\kappa_{\bot}$ is only $1\%-2\%$ of the parallel component $\kappa_{\|}$, despite the small development, vertical magnetic structures and outward flow arise, particularly in the exponential density model. 
The substantial increases in the velocity and altitude of the unstable flow reveal the importance of cosmic rays to the galactic wind.
In general, in the exponential density model, the morphology of the clumpy structure is filamentary. Conversely, in the hyperbolic tangent temperature model, a more concentrated and round shape of clumps like the giant molecular cloud are observed at the foot points of rising magnetic arches. The growth of instability in the hyperbolic tangent model is less rapid than that of the exponential density case since the pressure in the exponential density decreases faster with an increasing height.

The small gradient of cosmic ray pressure along the z direction caused by diffusion explains why an increase in $\kappa_{\|}$ facilitates the growth of the unstable undular mode. As a result, the gas falls down more rapidly, and adjacent loops join together to form a large loop-like bubble.

Exponential density with the adiabatic index $\gamma_{g}\sim 1$ (i.e. isothermal) produces the highest magnetic loops and most flow instability. A decreasing $\gamma_{g}$
leads to more condensed gas and, ultimately, a more unstable flow. While examining a situation without cosmic-ray diffusion, Parker indicated that the criterion for instability is $\gamma_{g}<1.36$. Although the undular mode is expected to be suppressed when $\gamma_{g}>1.4$, 
numerical results 
in our cases using $1.3<\gamma_{g}<1.4$  
indicate that the high adiabatic index still produces a non-uniform density distribution before the simulations end due to stringent numerical conditions. Therefore, the instability may still occur for large $\gamma$ if a stronger perturbation such as a supernova explosion is invoked.

The hyperbolic tangent temperature model yields a density distribution that increases with height, whose steepness depends on the width of the transition zone $w_{tr}$. For a smaller $w_{tr}$,  instability is increased as the density gradient correspondingly diminishes near the unstable regions.

When a larger ratio of plasma pressure to magnetic pressure $\alpha$ is applied,
the pressure inside the flux tube diminishes,
thus the instability growth is facilitated.
Also, as the length scale decreases accordingly, unstable structures are observed on a smaller scale.
Contrarily, a small $\alpha$ yields a larger scale height and so complicates the mixing of an unstable flow.

Finally, the ratio of plasma pressure to cosmic-ray pressure $\beta$ does not influence the length scale since the diffusion of the cosmic-ray pressure does not contribute to the scale height.

\section*{Acknowledgment}

We thank the National Center for High-Performance Computing Center for providing computing facilities 
and the National Science Council of Taiwan for funding this project under Grants 
NSC 98-2112-M-033-002, 
NSC 98-2811-M-033-014 and
NSC 98-2811-M-033-020.
CMK is supported by the Taiwan National Science Council Grants 
NSC 98-2923-M-008-01-MY3 and NSC 99-2112-M-008-015-MY3.
We are grateful to Jongsoo Kim for helpful comments.

\label{lastpage-01}

\end{document}